\documentclass[prc,showpacs,floatfix]{revtex4}
\usepackage{bm}
\usepackage{graphicx}
\usepackage{amsmath}
\usepackage{bbold}

\newcommand{\beq}{\begin{equation}}
\newcommand{\eeq}{\end{equation}}
\newcommand{\beqa}{\begin{eqnarray}}
\newcommand{\eeqa}{\end{eqnarray}}
\begin{document}

\title{
\hfill{\small {\bf MKPH-T-04-07}}\\
{\bf The generalized Gerasimov-Drell-Hearn sum rule for deuteron 
electrodisintegration}
}
\author{Hartmuth Arenh\"ovel}
\affiliation{Institut f\"{u}r Kernphysik, Johannes
Gutenberg-Universit\"{a}t, D-55099 Mainz, Germany}

\date{\today}

\begin{abstract}
The generalized Gerasimov-Drell-Hearn sum rule 
$I^{GDH}_{\gamma^*d}(Q^2)$ for deuteron 
electrodisintegration $d(e,e')np$ as function
of the  squared four-momentum transfer $Q^2$ 
is evaluated by explicit integration. The calculation is based on 
a conventional nonrelativistic framework using a realistic $NN$-potential 
and including contributions from meson exchange currents, 
isobar configurations and leading order relativistic terms. 
Good convergence is achieved. The prominent feature is a deep negative 
minimum, $I_{\gamma^* d}^{GDH}=-9.5$~mb, at low $Q^2\approx 0.2$~fm$^{-2}$
which is
almost exclusively driven by the nucleon isovector anomalous magnetic moment
contribution to the magnetic dipole transition to the $^1S_0$-state. Above 
$Q^2=20$~fm$^{-2}$ the integral $I^{GDH}_{\gamma^*}(Q^2)$ approaches 
zero rapidly.
\end{abstract}

\pacs{11.55.Hx, 24.70.+s, 25.30.Fj}

\maketitle

\section{Introduction}
\label{sec1}

The Gerasimov-Drell-Hearn (GDH) sum rule for real photons relates the 
square of the anomalous magnetic moment of a particle to the energy 
weighted integral $I^{GDH}_{\gamma}$ from threshold up to infinity over 
the beam-target spin asymmetry, i.e.\ the difference of the total 
photoabsorption cross sections for circularly polarized photons on a target 
with spin parallel and antiparallel to the spin of the photon, 
\begin{equation}
I^{GDH}_\gamma=4\pi^2\kappa^2\frac{e^2}{M^2}\,S
=\int_0^\infty \frac{d\omega^{lab}}{\omega^{lab}}
\left(\sigma_\gamma ^P(\omega^{lab})-\sigma_\gamma ^A(\omega^{lab})\right)
\,,\label{gdh}
\end{equation}
with mass $M$, charge $eQ$, anomalous magnetic moment $\kappa$ and spin $S$ 
of the particle. Furthermore, $\sigma_\gamma ^{P/A}(\omega^{lab})$ denote 
the total absorption cross sections for circularly polarized photons of energy
$\omega^{lab}$ on a target with spin parallel and antiparallel  
to the photon spin, respectively. The anomalous magnetic moment is 
defined by the total magnetic moment operator of the particle 
\begin{equation}
\vec M = (Q+\kappa)\frac{e}{M}\vec S\,.
\end{equation}

Previously, this sum rule has been evaluated for the deuteron 
by explicit integration up to an energy of 550 MeV including the contributions 
from the photodisintegration and single pion production
channels~\cite{ArK97,DaA03}. While for photodisintegration convergence 
was achieved yielding a negative contribution of $-413$ $\mu$b, 
the incoherent pion production contributions had not converged and a 
substantial positive contribution was still missing, as is needed to 
balance the negative result from photodisintegration in order to yield the 
small positive sum rule prediction  
$I^{GDH}_{\gamma,d} = 0.65\,\mu$b from the deuteron's small anomalous 
magnetic moment $\kappa_d=-0.143$.

It is the aim of the present paper to report on a first evaluation of the 
contribution of the electrodisintegration channel, i.e.\ $d(e,e')np$,
to the generalized 
GDH integral $I_{\gamma^* d}^{GDH}(Q^2)$ for the deuteron by explicit 
integration up to a maximum excitation energy of 1 GeV.

\section{The generalized GDH sum rule}
\label{sec3}

The spin asymmetry of the deuteron for real photons is related to 
the vector target asymmetry $\tau^c_{10}$ of the total photoabsorption
cross section~\cite{ArS91}, i.e., 
\begin{equation}
\sigma_\gamma ^P(\omega^{lab})-\sigma_\gamma ^A(\omega^{lab})=
\sqrt{6}\sigma_\gamma^0(\omega^{lab})
\tau^c_{10}(\omega^{lab})\,,\label{gdh_real} 
\end{equation}
where $\sigma_\gamma^0 $ denotes the unpolarized total photoabsorption cross
section. This spin asymmetry can be related to the transverse form factor 
$F^{\prime\, 10}_T$ of the inclusive electrodisintegration cross
section which appears for a vector polarized deuteron target in
conjunction with a longitudinally polarized electron beam. 

The general inclusive cross section for deuteron electrodisintegration 
including polarization degrees of freedom is 
governed by a set of ten inclusive form factors, namely two
longitudinal $F_L$ and $F_L ^{20}$, four transverse $F_T$, $F_T^{20}$,
$F_{TT}^{2-2}$, and $F^{\prime\, 10}_T$, and four 
longitudinal-transverse interference form factors 
$F^{1-1}_{LT}$,  $F_{LT} ^{2-1}$, $F^{\prime\, 1-1}_{LT},$
and $F^{\prime\, 2-1}_{LT},$ of which $F_{LT}^{1-1}$ and
$F^{\prime\, 2-1}_{LT}$ vanish below pion threshold due to time reversal 
invariance. Explicitly, the inclusive cross section reads~\cite{LeT91}
\begin{eqnarray}
\sigma_e (h, P^d_1 , P^d_2)&\equiv& 
\frac{d \sigma}{dk _2 ^{\mathrm{lab}} d \Omega _e ^{\mathrm{lab}}}\nonumber\\
&=& 6\,c(k_1^{\mathrm{lab}},\,k_2^{\mathrm{lab}})\, \bigl\{\rho _L F_L + 
\rho _T  F _T - P_1^d \rho_{LT} F_{LT}^{1-1} \sin \phi _d d_{10}^1
(\theta_d) \nonumber\\
& &{} + P _2 ^d \bigl[( \rho _L F _L ^{20} + \rho _T F _T ^{20}) d_
{00} ^{2} (\theta _d)- \rho _{LT} F _{LT} ^{2-1} \cos \phi _d d_{10} ^{2}
(\theta _d)
+ \rho _{TT} F _{TT} ^{2-2} \cos 2 \phi _d d _{20} ^{2}
(\theta _d) \bigr] \nonumber\\
& &{} + h P_1 ^d \bigl[ -\rho ' _T F^{\prime 10}_T d _{00}
^{1} (\theta _d) + \rho ' _{LT} F^{\prime 1-1}_{LT} \cos \phi _d d _
{10} ^{1} (\theta _d) \bigr]
- h P_2^d \rho '_{LT} F^{\prime 2-1}_{LT} 
\sin \phi _d d^2_{10}(\theta_d) \bigr\} \,, \label{incl_cross} 
\end{eqnarray}
where incoming and scattered electron momenta are denoted by
$k_1^{\mathrm{lab}}$ and $k_2^{\mathrm{lab}}$, respectively,
$c(k_1^{\mathrm{lab}},\,k_2^{\mathrm{lab}})$ and  
$\rho^{(\prime)}_\alpha$ ($\alpha\in\{L,\,T,\,LT,\,TT\}$) denote
kinematical factors, $h$ the degree of longitudinal electron polarization.
Furthermore, $P^d_{00}=1$, and $P^d_{1}$ and $P^d_{2}$ describe vector 
and tensor polarization of the deuteron, respectively, and the
spherical angles $(\theta_d,\phi_d)$ characterize the deuteron
orientation axis. The various form factors are functions of $E_{np}$,
the c.m.\ final state excitation energy, and of $q^{c.m.}$, the
three-momentum transfer in the c.m.\ system. 

At the photon point, $Q^2=(\vec q\,)^2-\omega^2=0$, the purely transverse form
factors are related to the various contributions of the general total 
photoabsorption cross section of deuteron photodisintegration, namely to 
the unpolarized total cross section $\sigma _\gamma^{tot}$ 
and to the beam and target asymmetries for polarized photons and
deuterons as defined in~\cite{ArS91}. In detail one has for $Q^2=0$
\begin{equation}
\sigma_\gamma ^{tot} = \frac{M_d}{{W_{np}\,q^{c.m.}}} F_T\,,\quad
\tau ^0 _{20} = \frac{F^{20} _T}{F_T}\,,\quad
\tau ^c _{10} = \frac{F^{\prime\, 10} _T}{ F_T}\,,\quad
\tau ^l _{22} = \frac{F^{2-2} _{TT}}{F_T}\,,
\end{equation}
where the invariant mass of the final $np$ system is denoted by
$W_{np}=E_{np}+2\,M$ with $M$ for the nucleon mass. 

Thus the spin asymmetry for real photons in (\ref{gdh_real})
corresponds to the vector target asymmetry for longitudinally polarized
electrons of the above inclusive cross section as defined by~\cite{LeT91} 
\beq
A_{ed}^V(\theta_d,\phi_d)=\frac{1}{4\,h\,P_1^d\sigma_e^0}
{\Big[}(\sigma_e(h,P_1^d,P_2^d)-\sigma_e(-h,P_1^d,P_2^d)
-\sigma_e(h,-P_1^d,P_2^d)+\sigma_e(-h,-P_1^d,P_2^d)\Big]\,,
\eeq
yielding for $(\theta_d,\phi_d)=(0,0)$, i.e.\ deuteron orientation
axis parallel to $\vec q$,
\beq
A_{ed}^V(0,0)=
\frac{6\,c(k_1^{\mathrm{lab}},\,k_2^{\mathrm{lab}})}{\sigma_e^0}
\rho'_T F^{\prime 10}_T\,,
\eeq
with $\sigma_e^0=\sigma_e(0,0,0)$ as unpolarized inclusive cross
section. 

Therefore, we introduce as spin asymmetry for transverse virtual photons
\beq
\sigma_{T,\gamma^*} ^P(\omega^{lab})-\sigma_{T,\gamma^*} ^A(\omega^{lab})=
\sqrt{6}\frac{M_d}{{W_{np}\,q^{c.m.}}}F^{\prime 10}_T
\,,\label{spin_asy_virtual} 
\eeq
which coincides at the photon point with Eq.~(\ref{gdh_real}).
Correspondingly, we take as extension of the GDH integral from real 
to virtual photons the definition~\cite{Dre01,DrT04}
\beq
I_{\gamma^* d}^{GDH}(Q^2)=\sqrt{6}\int_{\omega_{th}^{lab}}^\infty 
\frac{d\omega^{lab}}{\omega^{lab}}\,\frac{M_d}{W_{np}\,q^{c.m.}}
F_{T} ^{\prime 10}(E_{np},q^{c.m.})\,g(\omega^{lab},Q^2)\,,
\label{gdh_virtual}
\eeq
where $M_d$ denotes the deuteron mass. Here $E_{np}$ or equivalently 
$W_{np}=E_{np}+2\,M$ and $q^{c.m.}$ are functions of $\omega^{lab}$
and $Q^2$
\beq
W_{np}(\omega^{lab},Q^2)=\sqrt{M_d^2-Q^2+2M_d\,\omega^{lab}}\,\mbox{ and } 
q^{c.m.}(\omega^{lab},Q^2)=\frac{M_d}{W_{np}}\,\sqrt{Q^2+(\omega^{lab})^2}\,.
\eeq
The factor $g(\omega^{lab},Q^2)$ in (\ref{gdh_virtual}) takes into
account the fact, that the generalisation of the GDH integral is to a
certain extent arbitrary. The only restriction for this factor is the
condition that at the photon point $Q^2=0$ one has
\beq
g(\omega^{lab},0)=1\,,
\eeq
and that 
\beq
\lim_{\omega^{lab}\rightarrow \infty}g(\omega^{lab},Q^2)|_{Q^2=const.}<\infty
\eeq 
remains finite. As simplest extension we choose here 
$g(\omega^{lab},Q^2)\equiv 1$.

Transforming
(\ref{gdh_virtual}) into an integral over $E_{np}$ using
\beq
\omega^{lab}=\frac{1}{2M_d}\,(W_{np}^2+Q^2-M_d^2)
=\frac{1}{2M_d}\,((E_{np}+2M)^2+Q^2-M_d^2)\,,
\eeq
one obtains
\beq
I^{GDH}_{\gamma^* d}(Q^2)=
2\sqrt{6}\,M_d\int_0^\infty dE_{np}\frac{F_{T} ^{\prime
10}(E_{np},q^{c.m.})}
{(W_{np}^2+Q^2-M_d^2)\,q^{c.m.}}\,,
\label{gdh_virtual_a}
\eeq
where now $q^{c.m.}$ has to be considered as a function of $E_{np}$
and $Q^2$, i.e.\
\beq
q^{c.m.}(E_{np},Q^2)=\frac{1}{2W_{np}}\,\sqrt{((W_{np}-M_d)^2+Q^2)
((W_{np}+M_d)^2+Q^2)}\,.
\eeq

\section{Results for electrodisintegration}
\label{sec3a}

The generalized GDH integral of (\ref{gdh_virtual_a}) has been 
evaluated by explicit integration up to a maximum excitation energy 
$E_{np}=1$~GeV. The evaluation of $F^{\prime\,10}_T$ is
based on an expansion into transverse electric and magnetic multipole
matrix elements according to~\cite{LeT91}
\beqa
F_{T} ^{\prime 10}&=& 16 \pi^2  \sum_{LL'j \mu } (-)^{j}
\left(\begin{matrix}
L^{\prime}&L& 1 \cr 1 &-1&0 \cr
\end{matrix}\right)
\bigg\{ \begin{matrix}
L'&L&1 \cr 1&1&j \cr\end{matrix} \bigg\}e^{-2\rho_\mu^j}\, 
\Re e [(E^{L'}(\mu j)+M^{L'}(\mu j))^*(E^{L}(\mu j)+M^{L}(\mu j))]
\,,\label{ftprime}
\eeqa
where $\mu$ labels the possible final partial waves of given total angular 
momentum $j$ in the Blatt-Biedenharn parametrization~\cite{BlB52}, and 
$\rho_\mu^j$ its inelasticity which is zero below pion
threshold. Note, that due to parity conservation one has in
(\ref{ftprime}) either electric or magnetic contributions for a 
given multipolarity $L$ and state $\mu j$. 

The calculation is based on a nonrelativistic 
framework as is described in detail in
Ref.~\cite{ArS91,RiG97} but with inclusion of the leading order relativistic 
contributions. In the current operator we distinguish the one-body
currents with Siegert 
operators (N), explicit meson exchange contributions (MEC) beyond the
Siegert operators, essentially from $\pi$- and $\rho$-exchange, 
contributions from isobar configurations of the wave functions (IC), 
calculated either in the impulse approximation~\cite{WeA78} or in a
coupled channel approach for the most dominant
$N\Delta$-configuration~\cite{RiG97}, 
and leading order relativistic contributions
(RC). Bound and scattering states are obtained from a realistic
$NN$-potential for which we have chosen the Bonn r-space and q-space (B) 
models~\cite{MaH87} and the Argonne $V_{18}$ potential~\cite{WiS95}. 
The final state interaction (FSI) is taken into
account for all multipoles up to $L=6$ whereas for the higher
multipoles FSI can safely be neglected and plane waves are
used. 

\begin{figure}[htbp]
\includegraphics[width=.8\textwidth]{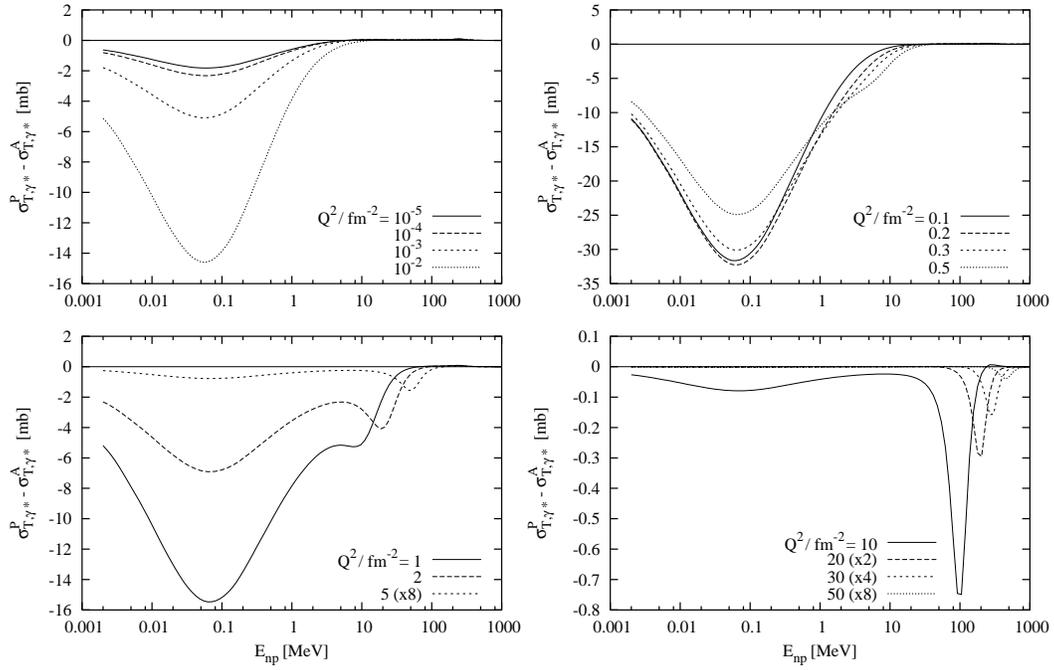}
\caption{Transverse spin asymmetry $\sigma_{T,\gamma^*}
^P-\sigma_{T,\gamma^*} ^A$ of deuteron electrodisintegration
$d(e,e')np$ as function of $E_{np}$ for various constant four-momentum
transfers $Q^2$. The calculation is based on the Argonne $V_{18}$ 
potential~\cite{WiS95} and includes all interaction and relativistic 
effects.
} 
\label{fig_gen_gdh1}
\end{figure} 
\begin{figure}[htbp]
\begin{center}
\includegraphics[width=.80\textwidth]{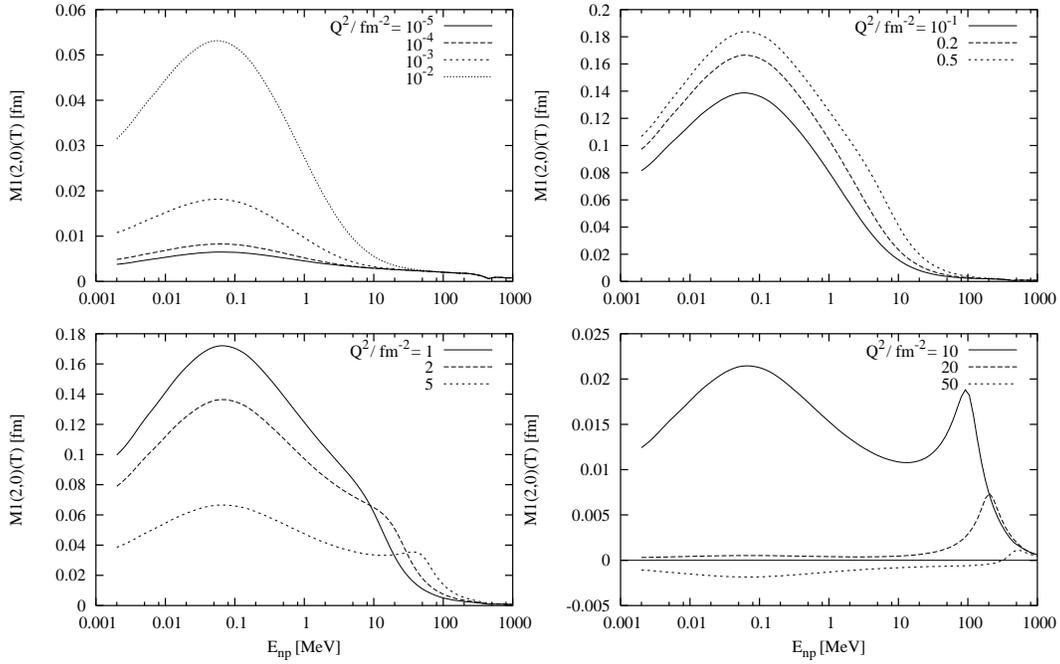}
\end{center}
\caption{Magnetic $M1(20)$-matrix element into the $^1S_0$-state 
for deuteron electrodisintegration 
$d(e,e')np$ as function of $E_{np}$ for various constant four-momentum
transfers $Q^2$ for Argonne $V_{18}$ potential~\cite{WiS95}. 
}
\label{fig_gen_gdh2}
\end{figure} 
In Fig.~\ref{fig_gen_gdh1} the transverse spin asymmetry
$\sigma_{T,\gamma^*} ^P- \sigma_{T,\gamma^*} ^A$ as
function of $E_{np}$ for various values of $Q^2$ are shown. The
prominent and most interesting feature, which one readily notes, is
the resonance like structure right above $np$-break-up threshold
around $E_{np}=70$~KeV. It stems essentially from the  
isovector $M1$-transition to the anti-bound $^1S_0$-state located at
this energy, which is well known from photo- and electrodisintegration
to dominate the cross section near threshold. Up to several MeV above
threshold the leading contributions come essentially alone from the
$L=1$-multipoles while 
the higher multipoles give a negligible contribution only. Restriction
to $L=1$ yields from (\ref{ftprime}) explicitly 
\beqa
 F_{T} ^{\prime 10}
&=&-\frac{8\pi^2}{3\sqrt{6}}\,\Big(2|M1(2,0)|^2+ |M^1 (1,1)|^2+ |M^1
(3,1)|^2- |M^1 (2,2)|^2- |M^1 (4,2)|^2\nonumber\\
&&+2|E1(3,0)|^2+ |E^1 (2,1)|^2+ |E^1 (4,1)|^2- 
|E^1 (1,2)|^2- |E^1 (3,2)|^2\Big)\,.
\eeqa
The $E1$-transitions leading to $^1P_1$ and $^3P_j$ ($j=0,1,2$) states
and which are most important in the inclusive cross section, do not play a
significant role in the spin asymmetry in this energy region. The reason 
for this feature is that the isoscalar transition to $^1P_1$ is 
largely suppressed, while the triplet $^3P_j$ contributions to
(\ref{ftprime}) almost cancel each other. The cancellation would be
complete if spin-orbit and tensor forces could be neglected, because
in this case the matrix elements are simply related by angular
momentum recoupling coefficients. Thus, at low energies only 
$M1$-transitions remain, essentially to $^1S_0$ and $^3S_1$ states. The
$^1S_0$ contribution is dominant because of the large isovector part
of the $M1$-operator arising from the large isovector anomalous
magnetic moment of the nucleon. It is particularly strong close to
break-up threshold at about 70 KeV, where the $^1S_0$ state is
resonant. This feature is seen in Fig.~\ref{fig_gen_gdh2} where this
matrix element is diplayed for various constant values of $Q^2$. Since this
state can only be reached by the antiparallel spin combination one
finds a strong negative spin asymmetry and thus a negative
contribution to the GDH integral. 
\begin{figure}[htbp]
\begin{center}
\includegraphics[width=.50\textwidth]{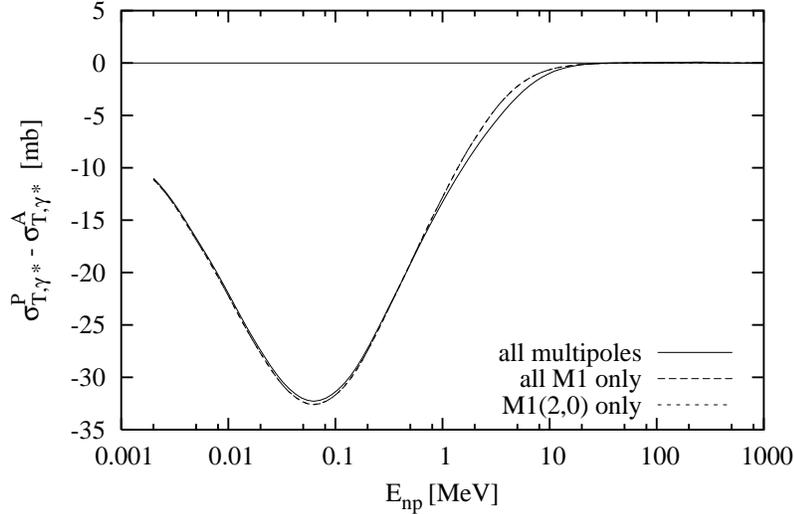}
\end{center}
\caption{Transverse spin asymmetry $\sigma_{T,\gamma^*}
^P-\sigma_{T,\gamma^*} ^A$ of deuteron electrodisintegration
$d(e,e')np$ as function of $E_{np}$ for $Q^2=0.2$~fm$^{-2}$
calculated including all multipoles (solid), all $M1$-multipoles only
(dashed) and the $M1$-transition into the $^1S_0$-state alone (dotted).
}
\label{fig_gen_gdh3}
\end{figure} 
The overwhelming predominance of the $M1$-transition into the $^1S_0$-state
is demonstrated in Fig.~\ref{fig_gen_gdh3} where a comparison of 
the spin asymmetry between calculations with all multipoles, with all 
$M1$-multipoles and with the $M1$-transition into the $^1S_0$-state alone
is displayed. The latter two coincide completely and also the calculation
including all multipoles shows only above $E_{np}\approx 1$~MeV a small
deviation. 

\begin{figure}[htbp]
\begin{center}
\includegraphics[width=.48\textwidth]{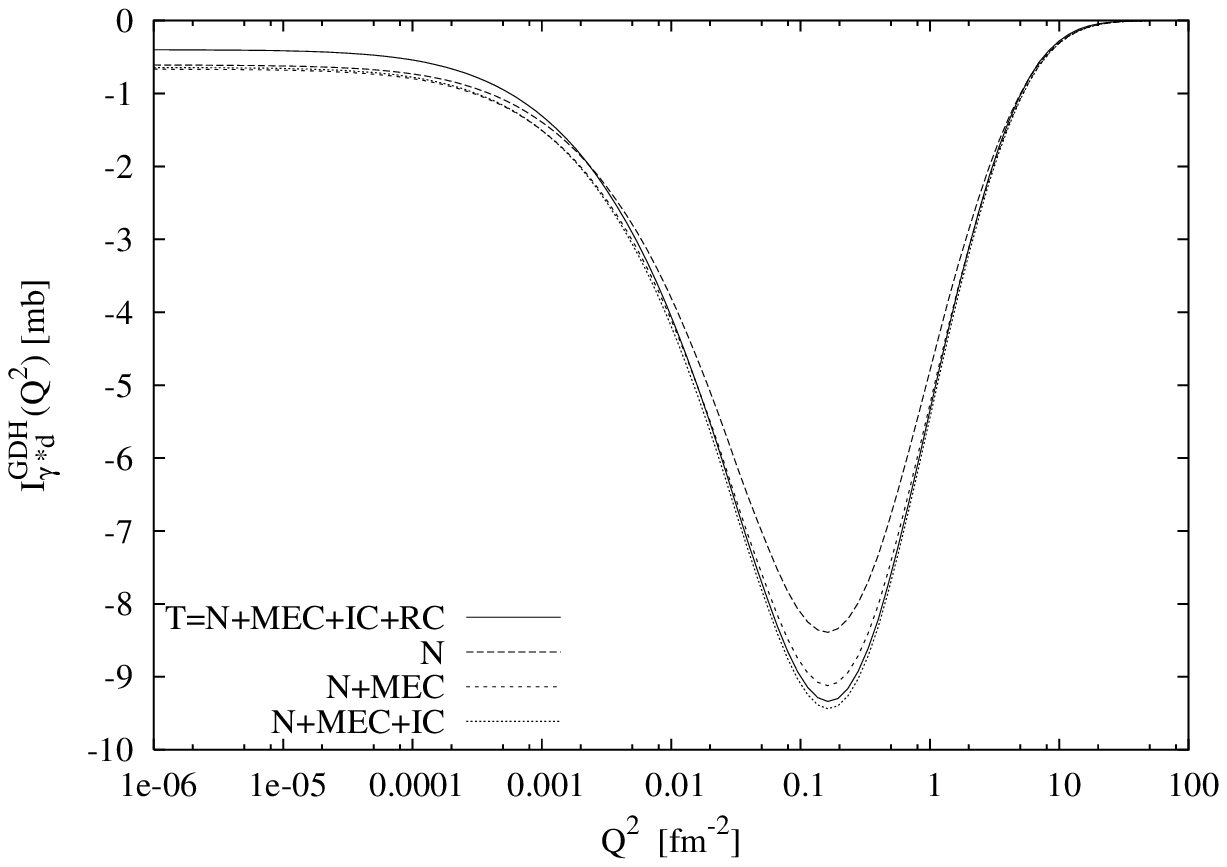}
\includegraphics[width=.48\textwidth]{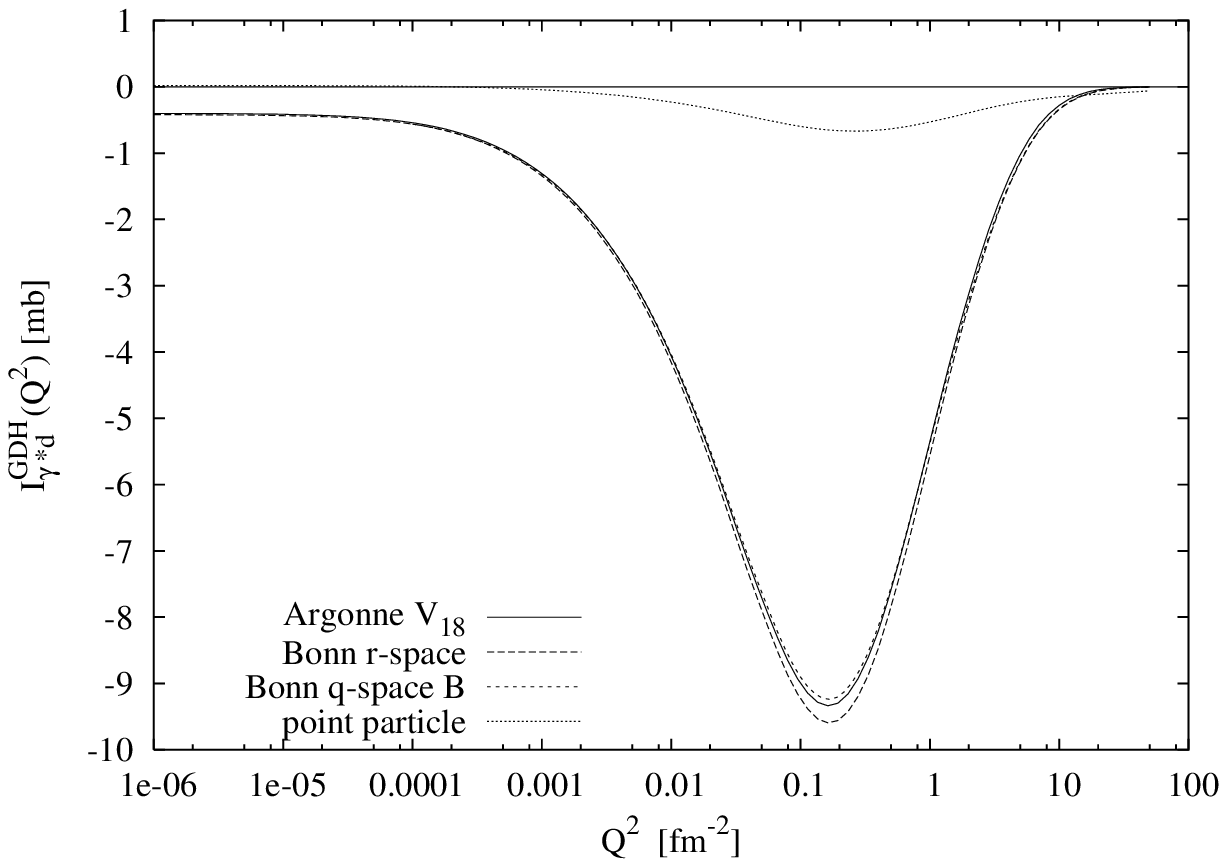}
\end{center}
\caption{Generalized Gerasimov-Drell-Hearn integral as function of
$Q^2$ for deuteron electrodisintegration $d(e,e')np$. Left panel:
separate current contributions from normal nonrelativistic theory (N)
and successively added meson exchange currents (MEC), isobar
configurations (IC), and relativistic contributions (RC). Right panel:
results of the complete calculation (T) 
for different potential models and for vanishing anomalous
nucleon magnetic moments (labeled ``point particle''). 
}
\label{fig_gen_gdh4}
\end{figure} 
Besides this low energy feature which, however, becomes less and less 
pronounced with increasing $Q^2$ above $Q^2=1$~fm$^{-2}$, one notes 
the evolution of the quasi-free peak as a distinct negative minimum 
in both the spin asymmetry as well 
as in the leading $M1$-matrix element located at $E_{np}/\mbox{MeV}\approx
10\,Q^2/\mbox{fm}^{-2}$ (see lower panels of Figs.~\ref{fig_gen_gdh1} 
and \ref{fig_gen_gdh2}). However, its size decreases rapidly with 
increasing $Q^2$. The rapid fall-off of the spin
asymmetry with increasing energy $E_{np}$ ensures furthermore that the
generalized GDH-integral converges sufficiently fast in view of the
additional energy weighting. In fact, convergence is achieved if one 
integrates up to an energy $E_{np}$ roughly 100~MeV above the quasi-free 
peak. 

The resulting $I_{\gamma^* d}^{GDH}(Q^2)$ is shown in
Fig.~\ref{fig_gen_gdh4}. For $Q^2\rightarrow 0$ the integral approaches 
$I^{GDH}_{\gamma d}$ for real photons. A pronounced minimum is 
readily seen around
$Q^2\approx 0.2$~fm$^{-2}$ reflecting the deepest minimum of the spin
asymmetries in Fig.~\ref{fig_gen_gdh1} for this value of $Q^2$. The
left panel shows the influence of the various interaction effects 
from MEC, IC and RC. Near the minimum, the
largest effect arises from MEC, increasing the depth by about 10~\%,
and to a smaller
extent from IC while their influences in the other regions of $Q^2$ is
quite small. Relativistic contributions are substantial near the
photon point as has been noted already for
photodisintegration~\cite{ArK97}. But at higher $Q^2$ they are quite
tiny. The right panel of Fig.~\ref{fig_gen_gdh4} shows a comparison of 
$I_{\gamma^* d}^{GDH}(Q^2)$ for three realistic potential models, the Bonn
r-space, the Bonn q-space (B)~\cite{MaH87} and the 
Argonne $V_{18}$~\cite{WiS95}
models. Obviously, the potential model variation is quite small
compared to the interaction effects. In view of the fact, that for
real photons $I_{\gamma,d}^{GDH}$ is driven by the nucleon anomalous magnetic
moments, we have also evaluated $I_{\gamma^* d}^{GDH}(Q^2)$ for vanishing
anomalous moments. The resulting integral, also shown in the right panel of
Fig.~\ref{fig_gen_gdh4}, is quite tiny, which underlines the fact 
that also the generalized GDH-integral is driven by the nucleon
anomalous magnetic moments. 

\section{Summary and conclusions}
\label{sec4}

The beam-target spin asymmetry of deuteron electrodisintegration for
transverse virtual photons and the associated generalized
Gerasimov-Drell-Hearn integral have been evaluated. The spin asymmetry
for constant four momentum transfer exhibits as function of the final
state excitation energy $E_{np}$ a very interesting low
energy property, a pronounced negative 
minimum around $E_{np}=70$~KeV, which is deepest for $Q^2\approx
0.2$~fm$^{-2}$. It is dominated by a single magnetic dipole transition
to the $^1S_0$-scattering state and almost completly governed by the
nucleon anomalous magnetic moment. All other multipoles play an
insignificant role. At higher excitation energies the spin asymmetry 
tends rapidly to zero, so that the generalized GDH-integral converges
fast, already at a few hundreds of MeV. The minimum in the spin asymmetry
leads to a corresponding negative minimum of $I_{\gamma^* d}^{GDH}(Q^2)$ around
$Q^2=0.2$~fm$^{-2}$. An experimental check of these predictions for
both the spin asymmetry as well as for the GDH-integral would provide
an additional significant test of our present understanding of low
energy behaviour of few-body nuclei. Furthermore, in view of this low energy
property, an independent evaluation in the framework of effective field
theory would be very interesting.

It remains as a task for future theoretical research to evaluate the spin
asymmetry and the GDH-integral for the other possible channels, like
coherent and incoherent single pion as well as two-pion electroproduction. 

\acknowledgments
I would like to thank Michael Schwamb for valuable discussions and a 
critical reading of the manuscript. 
This work was supported by the Deutsche Forschungsgemeinschaft (SFB 443).

\end{document}